\documentclass[conference]{IEEEtran}
\IEEEoverridecommandlockouts
\usepackage{cite}
\usepackage{amsmath,amssymb,amsfonts}
\usepackage{algorithmic}
\usepackage{graphicx}
\usepackage{textcomp}
\usepackage{xcolor}
\usepackage{algorithm}
\usepackage{multirow}
\usepackage{booktabs}
\usepackage{url} 
\usepackage{caption}

\usepackage{etoolbox}
\makeatletter
\patchcmd{\@makecaption}
  {\scshape} 
  {}         
  {}         
  {}         
\makeatother
\makeatletter
\patchcmd{\@makecaption}
  {\\}
  {.\ }
  {}
  {}
\makeatother
\definecolor{gain}{HTML}{2E7D32} 
\newcommand{\p}[1]{\textcolor{gain}{(#1)}}
\def\THD{large retailer}
\def\BibTeX{{\rm B\kern-.05em{\sc i\kern-.025em b}\kern-.08em
    T\kern-.1667em\lower.7ex\hbox{E}\kern-.125emX}}
\begin{document}

\title{Taxonomy-based Negative Sampling In Personalized Semantic
Search for E-commerce
}

\author{\IEEEauthorblockN{Uthman Jinadu}
\IEEEauthorblockA{\textit{Department of Computer Science} \\
\textit{Georgia State University}\\
Atlanta, Georgia, USA \\
ujinadu1@gsu.edu}
\and
\IEEEauthorblockN{ Siawpeng Er}
\IEEEauthorblockA{\textit{The Home Depot} \\
Atlanta, Georgia, USA \\
siawpeng\_er@homedepot.com}
\and
\IEEEauthorblockN{Le Yu}
\IEEEauthorblockA{\textit{The Home Depot} \\
Atlanta, Georgia, USA \\
le\_yu1@homedepot.com}
\and
\IEEEauthorblockN{Chen Liang}
\IEEEauthorblockA{\textit{The Home Depot} \\
Atlanta, Georgia, USA \\
chen\_liang@homedepot.com}
\and
\IEEEauthorblockN{Bingxin Li}
\IEEEauthorblockA{\textit{The Home Depot} \\
Atlanta, Georgia, USA \\
bingxin\_li@homedepot.com}
\and
\IEEEauthorblockN{Yi Ding}
\IEEEauthorblockA{\textit{Department of Computer Science} \\
\textit{Georgia State University}\\
Atlanta, Georgia, USA \\
yiding@gsu.edu}
\and
\IEEEauthorblockN{Aleksandar Velkoski}
\IEEEauthorblockA{\textit{The Home Depot} \\
Atlanta, Georgia, USA \\
aleksandar\_velkoski@homedepot.com}
}

\maketitle
\begin{center}
\vspace{-1.2em}
\textit{Accepted for publication at the 2025 IEEE International Conference on Big Data (IEEE BigData 2025).}
\vspace{0.8em}
\end{center}

\begin{abstract}
Large retail outlets offer products that may be domain-specific, and this requires having a model that can understand subtle differences in similar items. Sampling techniques used to train these models are most of the time, computationally expensive or logistically challenging. These models also do not factor in users' previous purchase patterns or behavior, thereby retrieving irrelevant items for them.
We present a semantic retrieval model for e-commerce search that embeds queries and products into a shared vector space and leverages a novel taxonomy-based hard-negative sampling(TB-HNS) strategy to mine contextually relevant yet challenging negatives. To further tailor retrievals, we incorporate user-level personalization by modeling each customer’s past purchase history and behavior. In offline experiments, our approach outperforms BM25, ANCE and leading neural baselines on Recall@K, while live A/B testing shows substantial uplifts in conversion rate, add-to-cart rate, and average order value. We also demonstrate that our taxonomy-driven negatives reduce training overhead and accelerate convergence, and we share practical lessons from deploying this system at scale.

\end{abstract}


\section{Introduction}
Online shopping has become an integral part of people's daily lives, making it crucial for e-commerce platforms to create high-quality, user-friendly, and efficient search engines. Delivering accurate and relevant product discovery is essential and challenging, directly impacting customer satisfaction and, ultimately, platform success. 

\begin{figure*}[!t]
     \centering
         \centering
         \includegraphics[width=.85\linewidth]{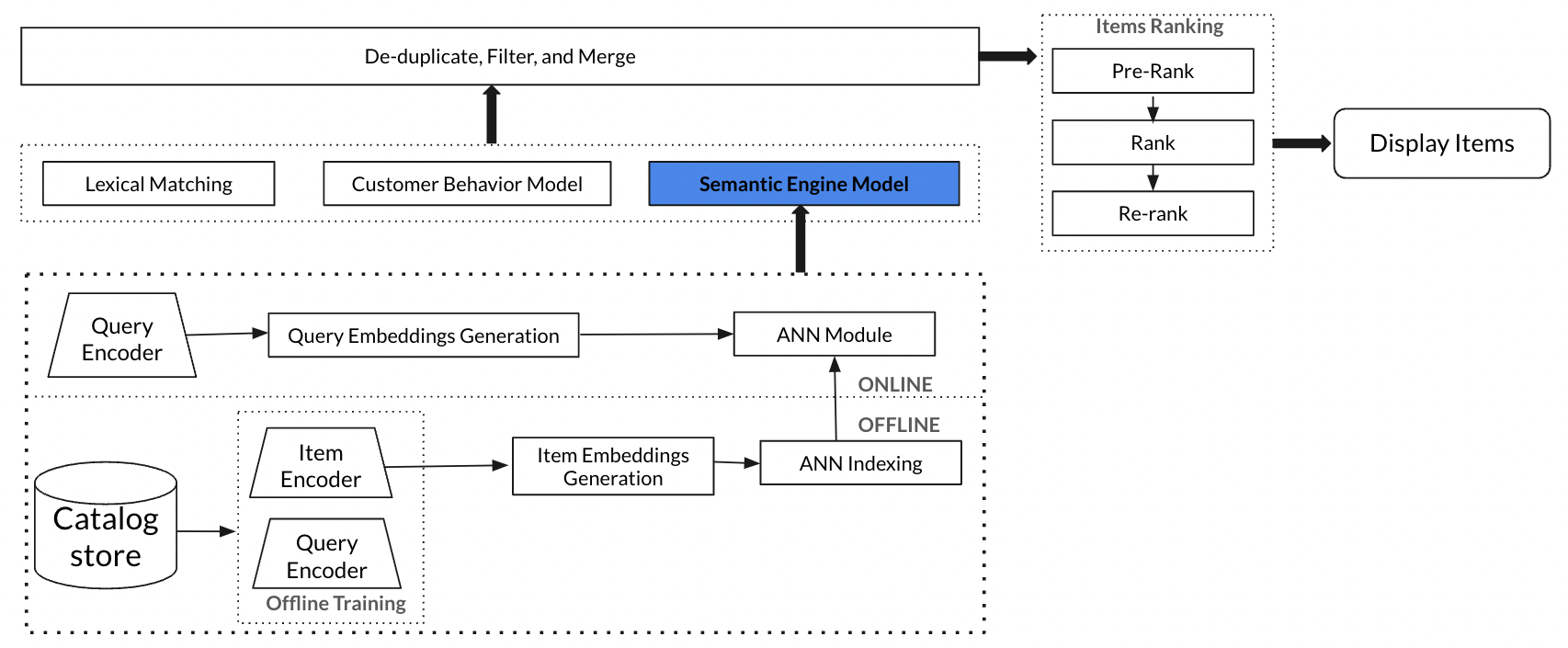} 
     \caption{Retrieval system architecture. \emph{Offline} (bottom), item metadata from the catalog are encoded by the item Encoder to produce item embeddings, which are indexed in an ANN service. \emph{Online} (top/bottom), a user query is encoded by the \textbf{Semantic Engine} to a query embedding, matched by the ANN module against the prebuilt index; candidates are then de-duplicated/filtered/merged and sent to a multi-stage ranking stack before display to the customer. The blue module marks the component we modify: the Semantic Engine. Our contribution is to train this model with our novel \emph{taxonomy-based hard-negative sampling}, enabling finer discrimination among closely related products. \emph{Personalized variant:} the Semantic Engine can fuse customer features and past purchases $(c,\;h_{\text{pur}})$ with the query via a dense layer to form a personalized query embedding $q_c$ before ANN retrieval; the ANN index and downstream ranking remain identical.}

     \label{fig:retrieval_architecture}
\end{figure*}

E-commerce search presents distinct challenges compared to web search. Text in e-commerce search is typically short and often unstructured, and leveraging extensive historical user behavior adds complexity. While lexical matching engines \cite{zobel2006inverted, duan2013supporting}, are valued for their reliability and precise control over search relevance, they fall short in bridging semantic gaps \cite{huang2020embedding}. Moreover, they struggle to account for user-specific preferences within the same query. 

The primary challenge for e-commerce platforms is to retrieve the most relevant products by effectively integrating query semantics with user behavior patterns.
Several companies have made significant strides in developing models for e-commerce applications, including Amazon \cite{xie2022embeddingbasedgrocerysearchmodel}, Amazon Search \cite{ai2019zero}, Walmart \cite{10.1145/3534678.3539164}, Microsoft \cite{10.1145/3539618.3591749}, and Taobao \cite{li2021embeddingbasedproductretrievaltaobao}, among others. Despite these impressive industrial deployments, most approaches still treat retrieval and personalization as separate problems, optimizing at scale but overlooking the relationship between a shopper's unique purchase history and the distinctions among similar products.

In reality, customers often exhibit varying purchasing patterns, from frequent, high-volume buyers to those with more occasional or specific needs. Additionally, the available products may have differences, requiring the semantic model to interpret and distinguish between similar items accurately. This complexity highlights the importance of having a model that could understand customer intent and connect it to the right products, ensuring a seamless and personalized shopping experience.

By leveraging signals such as purchase patterns and recent clicks, user engagement can be improved by delivering tailored and accurate items to individual customers.  We share insights from our work in developing a semantic retrieval model from the ground up, which addresses the challenges of capturing diverse customer behaviors and aligning them with relevant products. Although not every query requires personalization, our semantic model can detect when it is beneficial and fall back to a no-personalization mode otherwise. Figure \ref{fig:retrieval_architecture} presents a semantic retrieval system for e-commerce applications. This system consists of a number of industry-standard components. Our work focuses on the challenges of learning a semantic model to match query embeddings with a set of product embeddings. 


To optimize the semantic model, we need to use negative sampling due to the triplet loss. Products at the large retailer outlets may be domain-specific, which requires the semantic model to understand subtle differences in similar items. The current approach to sampling negatives is to apply random sampling, ANCE-style mining, or BM25-based sampling. However, these techniques have limitations. For example, randomly selecting negative examples often yields items that are completely unrelated to the target product, offering semantically irrelevant items. Additionally, existing negative sampling methods work well for small and well defined problems, or may not be suitable for our application. 


To address this, we present a taxonomy-based hard-negative sampler(TB-HNS) for e-commerce search. Our taxonomy-based sampler leverages category hierarchies to select negatives that are semantically related yet irrelevant: by moving up one level in the taxonomy, we pick items sharing broader categorical context but distinct semantics, producing more informative hard negatives and boosting training efficiency.

Our main contributions to the state-of-the-art retrieval systems are as follows:
\begin{enumerate}

    \item We developed a semantic model that can distinguish between closely related but irrelevant items through a novel taxonomy-based hard-negative sampling generation method, ensuring more precise and relevant search outcomes.
    
    \item We integrate personalization, modeling each customer’s past interactions and preferences, and effectively handle cold-start items, ensuring newly added products are accurately retrieved

    \item We demonstrate that our model effectively aligns generic and less specific queries from infrequent shoppers, improving retrieval accuracy through enhanced personalization and synonym mapping.

    \item Our taxonomy-based hard-negative sampling(TB-HNS) not only outperformed random, ANCE, and BM25‐based negatives in recall, retrieval relevance, and query–item alignment, but also simplified training by reducing negative sampling overhead and improving data preparation efficiency.

   \item We demonstrate that our taxonomy-based hard negative sampling transfers beyond the home-improvement domain: applied to the public Amazon ESCI dataset \emph{without dataset-specific tuning} and under the same training/evaluation protocol, it consistently outperforms random, BM25, and ANCE mining across Recall@k while retaining its latency advantages.

    \item We share practical lessons from deploying our system at scale on a platform serving millions of daily customers.

\end{enumerate}
\section{Related Works}
\begin{figure*}[!t]
     \centering
         \centering
         \includegraphics[width=1.0\linewidth]{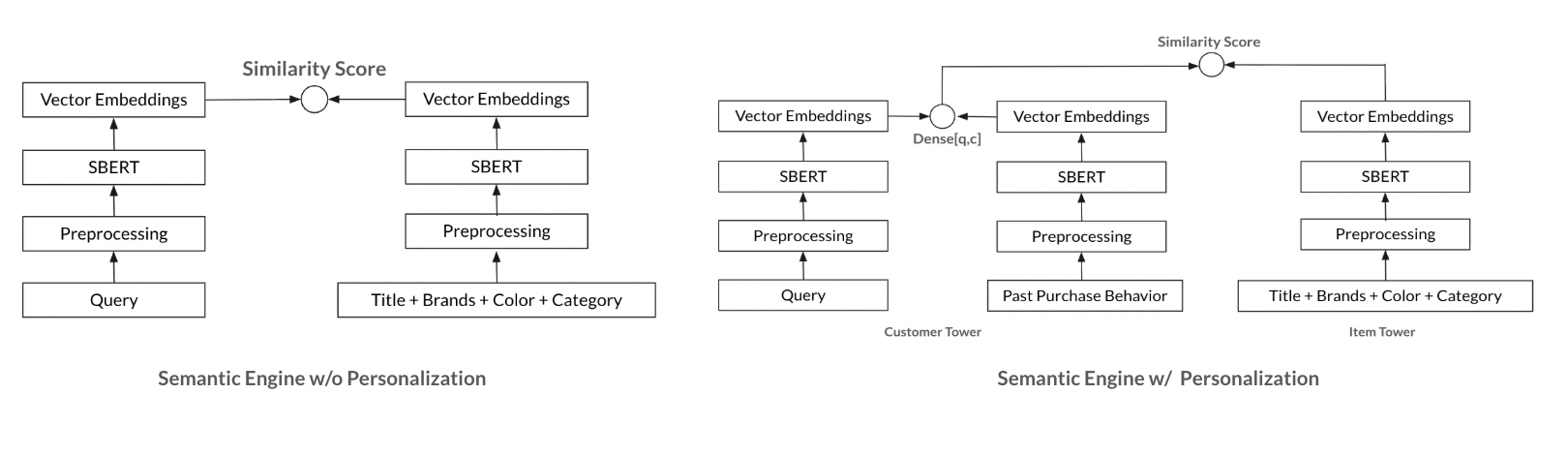} 
     \caption{Semantic Engine without personalization (left): a two-tower bi-encoder where the query and item are encoded and scored with a dot product. Enhanced Personalized Semantic Engine (right): augments the baseline with a customer tower that fuses the query with profile features \(c\) and purchase history \(h_{\text{pur}}\) via a dense layer to form a personalized query embedding \(q_c\); the item tower and similarity function remain unchanged.}

    \label{fig:models}
\end{figure*}

\textbf{\emph{Semantic Search Approaches:}}
Two-tower models, also known as dual encoders or Siamese networks, have become a popular choice in embedding-based neural systems across a wide range of applications, including passage/document retrieval \cite{xiong2020approximatenearestneighbornegative, chang2020pretrainingtasksembeddingbasedlargescale}, recommender systems \cite{10.1145/3366424.3386195, 10.1145/3616855.3635841}, and dialogue systems \cite{mazare-etal-2018-training}. 
Two-tower architectures often incur prohibitive complexity and latency for real-time, large-scale use. We tackle this by engineering an e-commerce retrieval model that delivers high accuracy with low-latency performance.

\noindent\textbf{\emph{Embedding-driven Retrieval Systems:}}
Nigam et al. introduced semantic embedding retrieval \cite{nigam2019semantic}; Wu et al. proposed zero-shot techniques \cite{wu2020zero}; Facebook described full-stack embedding optimizations \cite{liu2021que2search,huang2020embedding}; Taobao and JD developed personalized product search systems \cite{li2021embeddingbasedproductretrievaltaobao,zheng2022multiobjectivepersonalizedproductretrieval}. Amazon used tree-based extreme multi-label classification and “Zero Attention” personalization \cite{chang2021extreme,ai2019zero}; Instacart applied a two-tower transformer for query–product embeddings \cite{xie2022embeddingbasedgrocerysearchmodel}; Walmart combined inverted indexing with neural embeddings for long-tail queries \cite{magnani2022semantic}. We extend these approaches by jointly optimizing for semantic relevance and personalization.

\noindent\textbf{\emph{Strategies for Negative Item selection:}}
In‐batch negative sampling \cite{yih2011learning} reuses mini‐batch examples as negatives, avoiding explicit labels and reducing compute. Streaming caches \cite{he2020momentum} and hybrids of in‐batch random with offline hard negatives \cite{magnani2022semantic} diversify contrastive signals. Yet iterative ANCE mining \cite{xiong2020approximatenearestneighbornegative} is impractical when queries match hundreds of products, and BM25 sampling \cite{karpukhin2020dense} yields many false negatives and high indexing costs on terse queries. We adopt a triplet‐loss framework with taxonomy‐based negatives to generate semantically tough yet scalable training pairs.

\noindent\textbf{\emph{Personalization of Search:}}
Personalization re‐ranks results using signals like location, history, and clicks \cite{dumais2016personalized, ai2019zero, chirita2005using, shen2005implicit}, yet identical queries can trigger diverse behaviors and uneven gains \cite{dou2007large, chuklin2022click}. Jannach and Ludewig \cite{jannach2017investigating} cast product‐search personalization purely as recommendation, discarding query context. We extend these methods by integrating multiple positives, taxonomy‐based hard negatives, and adaptive personalization to balance semantic relevance with user signals for more precise, tailored retrieval.



\section{Model Definition}
\begin{figure*}[!t]
    \centering
    \includegraphics[width=.60\linewidth]{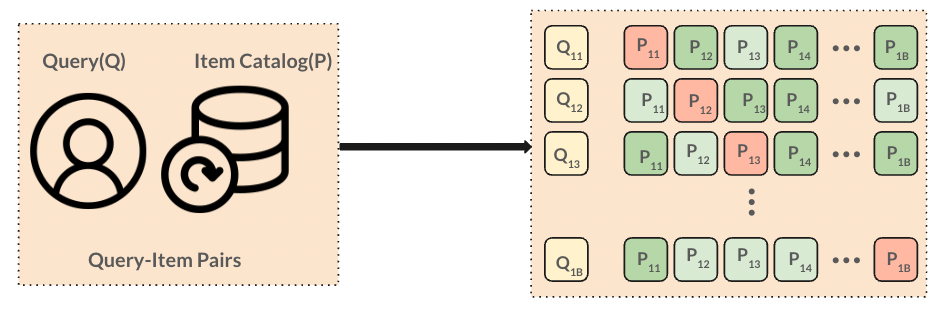}
    \caption{Taxonomy-based hard-negative sampling. For each query $Q_i$, the ground-truth positive item $P_i^\star$ is shown in \textcolor{red}{red}. Candidate hard negatives $P_{i,m}$ are shown in \textcolor{green!60!black}{green}, with different shades indicating different hard-negative items. Negatives are sampled from sibling items under the same parent category in the product taxonomy as $P_i^\star$. This yields ``near-miss'' negatives-contextually similar but not identical to the positive, so the model learns fine-grained distinctions (e.g., brand, size, finish) rather than relying on broad category features.}
    \label{fig:hard-negative-generation}
    \vspace{-.35cm}
\end{figure*}

We first review the legacy Customer Behavior and Lexical Matching models, outlining their limitations, and motivate a new approach. We then formalize the task and present two semantic retrieval methods: a non‐personalized baseline and our enhanced, purchase‐aware semantic engine.

\subsection{Existing Retrieval System and its limitations at the Large Retailer}
\subsubsection{Lexical Matching Model:} 
The Lexical Matching Model is a lightweight, exact-match retrieval system that scores products by combining keyword overlap with a document-quality signal via tunable weights. Its simplicity and speed make it highly effective when queries use the same terms found in titles or descriptions. However, because it relies solely on exact keyword matches, it breaks down on ambiguous or synonym-rich queries, failing to bridge the semantic gap when users’ wording doesn’t align verbatim with the product catalog.

\subsubsection{Customer Behavior Model:} 

The Customer Behavior Model is a keyword-based model that accounts for historical user behavior (e.g. clicks and other item interaction events) to capture unique customer properties such as terminology use. This is important because customers may refer to a small seating area as a 'cozy nook'. However, because the model is matching-based, 'cozy nook' may not refer to accent chairs or small seating furniture that the customer is interested in purchasing. Additionally, the model often overlooks newly cataloged or less popular products, as well as important item attributes.


While these approaches initially met most e-commerce system needs, it became apparent that the retrieval system could not provide the ranking system with sufficiently suitable candidates. 
Moreover, matching-based approaches fail to account for individual shopping behaviors and expectations effectively, and therefore requires an improved personalization method. We now discuss how we address these challenges using a semantic retrieval model.

\subsection{Developing a Semantic Engine Model for Retrieval}

We develop our semantic engine model motivated by existing understanding of two-tower based approaches. Facebook outlined comprehensive embedding optimization strategies \cite{liu2021que2search,huang2020embedding}, focusing on enhancing search and recommendation systems.
Instacart utilized a two-tower transformer for query–product embedding generation \cite{xie2022embeddingbasedgrocerysearchmodel}, a dual-encoder architecture that separately processes queries and products, enabling efficient matching and ranking in e-commerce search. 

Let \( U \) represent the set of customers, \( Q \) denote the corresponding customer queries, \( I \) stand for the collection of items in the {\THD} item catalog, and \( C \) represent customer profiles. We also aggregate each customer’s past purchase history over a recent period leading up to the current purchase.

Given the historical purchase behavior of customer \( C \), the task is to return a set of items \( i \in I \) that match the customer’s query \( q \), submitted at time \( t \). Specifically, the goal is to predict the top-\( K \) item candidates from \( I \) at time \( t \) based on a score \( z \), which measures the relevance between the customer's information (query, past purchases) and the items. From this setup, we establish two major model formulations: a baseline model that excludes personalization based on customer-specific data, and an enhanced personalization model that incorporates customer-specific features. The formal definitions of these models are as follows:

\subsubsection{Semantic Engine Model}
\label{subsec:baseline_model}
Our semantic engine model employs a two-tower bi-encoder architecture optimized for semantic search, and can formally be defined as:

\begin{equation}
t = S \left( \theta(q), \eta(i) \right)
\label{eq:eqn_baseline_model}
\end{equation}
where \( S(\cdot) \) is the scoring function, \( \theta(\cdot) \) encodes the query, and \( \eta(\cdot) \) encodes the item. We rely on a two-tower retrieval model, where the scoring function \( S \) is an inner product, \( \theta \) and \( \eta \) is a BERT-based model implemented within the Sentence-BERT(SBERT) \footnote{\url{www.sbert.net}}  framework. Please see figure \ref{fig:models}.

\paragraph{Query Tower}
 We have a query tower that takes the search term entered by customers into the search bar as input. The term, representing the item customers wish to purchase, is processed through the model to generate embeddings, which we refer to as the query-embeddings.

\paragraph{Item Tower}
The item tower processes metadata from the items catalog, which includes details such as the item's title, brand, color, and other relevant attributes. These metadata are fed into the model, which produces embeddings representing each item. These item embeddings capture the essential characteristics of the items and enable the search system to match them effectively with queries and customer-specific data. 

\subsubsection{Enhanced Personalization Search Model}
\label{subsec:enhanced_personalization}
To tailor search results through personalization, we extend the model by introducing a customer-specific tower, which integrates both past purchase behaviors and demographic features into the search process with \( h\_pur \) and \( c \) combined with the query tower via a dense layer. This enhanced model is defined as:

\begin{equation}
t_c = S \left( \theta(q, c, h\_pur), \eta(i) \right)
\label{eq:eqn_enhanced_model}
\end{equation}

where \( c \) represents the customer’s demographic features, and \( h\_pur \) refers to the customer's past purchase behavior. In this case, \( \theta(\cdot) \) encodes the query, customer demographics \( c \), and past purchases \( h\_pur \), while \( \eta(\cdot) \) encodes the item. The model again adopts a two-tower architecture, but now with the query tower fused via a dense layer with the customer features and past purchase behavior, and \( S \) is still instantiated using the inner product function for efficient computation. Please see figure \ref{fig:models}

\textit{Customer Tower:}
In our enhanced personalization model, we have an additional tower,  the customer tower, that combines the query \( q \) with the user profile \( c \). This fusion of the search query and customer-specific data results in a combined embedding, denoted by \( q_c \). This embedding incorporates both the query and personalization aspects, allowing the model to tailor retrievals to individual customer preferences.

\subsection{Loss Function}
We use Multiple Negative Ranking Loss (MRNL) to train our model. This objective function requires \texttt{[query, positive, negative]} triplets. MRNL works by minimizing the distance between query-positive embeddings while maximizing the distance from negative samples. A high similarity score for the positive pair reduces the loss, while similar scores for negatives increase the loss. The loss is defined as:
\begin{equation}
\mathcal{L} = - \log \left( \frac{\exp(\text{sim}(q, p))}{\exp(\text{sim}(q, p)) + \sum_{i=1}^{|N|} \exp(\text{sim}(q, N_i))} \right)
\label{eq:eqn_multinegativerankingloss}
\end{equation}
Here, $sim(q, p)$ refers to the similarity score (inner product) between the query \(q\) and the positive example \(p\). $N$ is a set of negative example embeddings. 

\subsection{Taxonomy-based Hard Negative Sampling}

\begin{figure}[t]
    \centering
    \begin{algorithm}[H]
        \caption{Taxonomy-Based Hard Negative Sampling}
        \label{alg:taxonomy_hard_negative_sampling}
        \begin{algorithmic}[1]
            \REQUIRE Query item \( Q \), Positive set \( P \), Maximum attempts \( K \)
            \ENSURE Hard negative sample \( N \) or \texttt{None}
            
            \STATE Extract the taxonomy \( \mathcal{T}(Q) \) of the query item.
            \STATE Identify the item's parent category in the taxonomy:
            \[
            \mathcal{T}_p(Q) = \text{Parent}(\mathcal{T}(Q))
            \]
            \STATE Retrieve the set of candidate items under the parent category:
            \[
            \mathcal{C} = \{ C_i \mid C_i \in \mathcal{T}_p(Q) \}
            \]
            \STATE Initialize attempt counter \( i \leftarrow 0 \).
            \WHILE{$i < K$}
                \STATE Sample \( N \) uniformly at random from \( \mathcal{C} \).
                \IF{$N \notin P$}
                    \RETURN \( N \)
                \ENDIF
                \STATE Increment \( i \leftarrow i + 1 \).
            \ENDWHILE
            \RETURN \texttt{None}
        \end{algorithmic}
    \end{algorithm}
    \vspace{-.75cm}
\end{figure}

Due to unique data characteristics, we found that existing methods for negative sampling did not work well for the triplet loss. We therefore propose a novel taxonomy-based hard negative sampling method presented in Algorithm \ref{alg:taxonomy_hard_negative_sampling}.

Random negative sampling \cite{breiman2001random} often selects semantically irrelevant items, failing to provide challenging negatives for robust learning. ANCE \cite{xiong2020approximatenearestneighbornegative} iteratively mines hard negatives but struggles with scalability when hundreds of relevant products exist per query, becoming computationally expensive. Similarly, BM25-based sampling \cite{karpukhin2020dense} struggles with short, ambiguous e-commerce queries, leading to false negatives and requiring extensive indexing and scoring over large catalogs, which is computationally costly and misses domain-specific details like brand or color.

We propose a novel technique called taxonomy-based hard negative sampling, designed to enhance the model's ability to retrieve relevant items and improve recall metrics. Unlike traditional sampling methods, which assume a predefined set of relevant items, our approach accommodates scenarios where numerous relevant items may exist for a single query. For instance, a query such as "Moving Boxes" might correspond to thousands of relevant items. To tackle the challenge of evaluating such queries, we introduce heuristics to generate hard negatives.

Our approach leverages the hierarchical taxonomy of items, which organizes them into categories and subcategories. First, we identify the parent category to which a positive item belongs. Then, by moving to its immediate higher-level category, we retrieve candidate items for generating hard negative samples (see Figure \ref{fig:hard-negative-generation}). This strategy ensures that the negatives are contextually similar yet distinct from the query.  

To avoid incorrect associations, any candidate overlapping with the positive set is excluded. If no valid candidate is found after multiple attempts, no negative sample is returned for that iteration. The process is in Algorithm \ref{alg:taxonomy_hard_negative_sampling}.

\section{Integration into Production System}

We discuss how we leverage our trained semantic engine model to power both the retrieval and ranking stages of our e-commerce system. To efficiently retrieve similar items, we utilize Facebook AI Similarity Search (FAISS) \cite{johnson2017billionscalesimilaritysearchgpus}, a library for Approximate Nearest Neighbor (ANN) search. FAISS indexes high-dimensional item embeddings using techniques like clustering and optimized quantization to approximate nearest neighbors of a query embedding. This approach reduces computational complexity and enables efficient retrieval of large-scale datasets.

Once relevant items are retrieved, their similarity scores, calculated based on the distance between the query embedding and the item embeddings, are fed into the ranking system. The similarity scores are then used to sort retrieved results. 


\subsection{Predeployment Training and Index setup}
We begin by leveraging items purchased engagement data to train an embedding model using Sentence-BERT (SBERT) framework. This trained model converts textual information about items (such as title, brand) into dense vector representations, ensuring that semantically similar products are mapped close to each other in this high-dimensional space.

Once we generate these embeddings for all items, the next step is to efficiently index and retrieve them. We employ FAISS to build an Approximate Nearest Neighbor (ANN) index for fast retrieval. 


\subsection{Online Search and Retrieval}
The system transforms a user's query into an embedding vector using the finetuned model, which is then matched against item embeddings stored in the FAISS index. This retrieves the top-K most similar items based on dot product, ensuring fast and relevant results. 
Once the most relevant items are retrieved, the system filters out items with low similarity scores and confirms that only in-stock items are displayed, guaranteeing both relevance and availability.
\section{Experiments Setup}
In this section, we discuss how we setup experiments to evaluate our model and  taxonomy-based hard negative sampling. We discuss datasets, metrics, preprocessing and implementation details.


\begin{table*}[t]
\caption{Recall@24 and Recall@100 for two standard baselines (DistilledBERT \cite{Hofstaetter2021_tasb_dense_retrieval}, BM25 \cite{trotman2014improvements}) and our three semantic engine variants: non-personalized, personalized, and combined. Personalization substantially improves recall, and the Combined Model achieves the highest performance, ensuring robust retrieval even when user history is unavailable. The combined model is trained using both the personalization and non-personalization components.}
\centering
\small
\begin{tabular}{lrrrr}
\toprule
\textbf{Model} & \textbf{Recall@8} & \textbf{Recall@12} & \textbf{Recall@24} & \textbf{Recall@100} \\
\midrule
DistilledBert \cite{Hofstaetter2021_tasb_dense_retrieval} & 41.68 & 40.21 & 36.39 & 39.51 \\
BM25 \cite{trotman2014improvements} & 17.02 & 21.62 & 30.94 & 49.30 \\
\midrule
Ours w/o Personalization & 52.85 & 59.93 & 68.51 & 78.34 \\
Ours w/ Personalization & 63.5 & 69.88 & 77.17 & 83.66 \\
Combined Model(Ours) & \textbf{63.86} &  \textbf{69.90} & \textbf{77.89} & \textbf{84.23} \\
\bottomrule
\end{tabular}
\label{tab:recall_model_result}
\vspace{-.30cm}
\end{table*}

\subsection{Baseline Sampling Methods}

We compare against the following negative sampling techniques.

\textbf{Random negative }sampling \cite{breiman2001random} involves selecting negative examples arbitrarily, often resulting in unrelated items that are semantically irrelevant. Though straightforward to implement, it frequently fails to provide the model with challenging negatives necessary for robust learning, particularly in an e-commerce setting where distinguishing subtle relevance is key for query-product matching. 

Xiong et al. \cite{xiong2020approximatenearestneighbornegative} proposed Approximate Nearest Neighbor Negative Contrastive Learning (\textbf{ANCE}), an iterative hard-negative mining procedure that, at each step, identifies and incorporates difficult negatives into the training set. While this method works well when the set of “relevant” items is small and well-defined, in a broad e-commerce context there may be hundreds of genuinely relevant products per query, making their approach not only logistically challenging but also computationally expensive to run at scale.

Karpukhin et al \cite{karpukhin2020dense} utilized \textbf{BM25} for hard negative sampling in Dense Passage Retrieval, selecting top-ranked but non-positive passages to enhance contrastive learning. While effective for question answering, BM25 struggles in e-commerce due to short , ambiguous queries often leading to false negatives generated. Additionally, computing BM25 scores over large e-commerce catalogs is computationally expensive, requiring extensive indexing and scoring. Another issue is that in e-commerce retrieval where catalogs contain millions of products described by domain-specific attributes (e.g brand, color etc) and require semantic matching, BM25-based negatives often fail to capture these nuances, resulting in poorer performance.

            

\subsection{Evaluation Metrics}
\subsubsection{Offline Evaluation Metrics}

To assess offline retrieval, we measure \textit{Recall@K}. Let \(T=\{t_1,\dots,t_N\}\) be the set of true relevant items and \(I=\{i_1,\dots,i_K\}\) the model’s top-\(K\) predictions. Then
\[
\text{Recall@K} \;=\;\frac{|I \cap T|}{|T|}\,,
\]
which quantifies the fraction of relevant items recovered within the top-\(K\) results.




\subsubsection{Online Evaluation Metrics}
We assess live performance via A/B testing using four key metrics: Conversion Rate (CR), the fraction of sessions ending in $\ge 1$ purchase; Add‐to‐Cart Rate (ATC), the fraction of sessions with $\ge 1$ add‐to‐cart action; Average Order Value (AOV), the mean revenue per order; and 95th‐percentile latency (P95), defined as the smallest \(L_{95}\) satisfying
\[
\Pr(\text{latency} \le L_{95}) \ge 0.95,
\]
to ensure our service‐level objectives under peak load.  

\subsection{Datasets}
\label{sec: Datasets}
We trained our model on a 24-month window of the large retailer customer engagement, producing roughly 4 million personalized and 4 million non-personalized triplets, and evaluated it on a held-out 4-month period comprising about 6 thousand examples in each setting. During preprocessing, we removed duplicates and empty entries, aggregated behaviors by query and customer ID (clicks, add-to-cart events, purchases), and built a catalog lookup keyed by item ID for fast access to product metadata.

Both training and testing data are provided in two variants:
\begin{itemize}
  \item \textbf{Personalized} (PER\_Train, PER\_Test): includes past purchases and customer-specific features.
  \item \textbf{Non-personalized} (NPER\_Train, NPER\_Test): omits all customer history and personal attributes.
\end{itemize}
For example, a personalized data sample contains information in the following format: 
\[
[\text{query} + \text{past\_purchase} + \\\text{customer\_info},\;\text{positive},\;\text{negative}\;]
\]
Non-personalized ones only contains [\;\text{query},\;\text{positive},\;\text{negative}\;]
with “positive” denoting purchased items and “negative” drawn via taxonomy-based hard-negative sampling. 

To assess generalization beyond the home-improvement domain, we evaluated TB-HNS on the public Amazon ESCI dataset~\cite{reddy2022shoppingqueriesdatasetlargescale} under the same training/evaluation protocol, confirming effectiveness across broader e-commerce categories.



\subsection{Implementation Details}
For our experiments, we used the msmarco-distilbert-tas-b \cite{Hofstaetter2021_tasb_dense_retrieval} model as the pretrained base, selecting it for its lightweight architecture, speed, and performance parity with BERT, while reducing latency during production. 


\begin{table*}[t]
\caption{Recall performance comparison of sampling techniques. Our \textit{taxonomy-based hard negatives} significantly outperform other negative sampling methods in Recall@k = (8, 12, 24, and 100) metrics, with \textbf{38--48\%} relative gains over the strongest non-taxonomic baseline (ANCE) and consistent improvements from early to deep ranks. 
Bold numbers indicate the best result per column.}
\centering
\small
\begin{tabular}{lrrrr}
\hline
\textbf{Sampling Technique} & \textbf{Recall@8} & \textbf{Recall@12} & \textbf{Recall@24} & \textbf{Recall@100} \\
\hline
Random Negative \cite{breiman2001random} & 41.89 & 44.53 & 47.97 & 51.96 \\
Karpukhin et al (BM25) \cite{karpukhin2020dense}  & 38.37 & 40.89 & 44.25 & 48.97 \\
Xiong et al (ANCE) \cite{xiong2020approximatenearestneighbornegative} & 45.96 & 48.43  & 52.32 & 60.43 \\
Taxonomy-based Negative (Ours) & \textbf{63.50} & \textbf{69.88} & \textbf{77.17} & \textbf{83.66} \\
\hline
\end{tabular}
\label{tab:compare-negative-sampling-techniques}
\end{table*}

We fine-tuned our model using Sentence-BERT framework which utilizes a siamese network architecture and contrastive learning objectives to generate sentence embeddings.  We optimized hyperparameters through a validation dataset to achieve stable and effective performance. The training process was conducted on a high-performance computing platform to ensure efficiency and scalability. Additionally, our training and evaluation setup accounted for model deployment considerations such as latency and other production requirements.

\section{Results}
\subsection{Offline Evaluation Result}
We begin by presenting the evaluation of our baseline model without personalization, as detailed in Section \ref{subsec:baseline_model}. We then report the performance of our enhanced personalization model (Section \ref{subsec:enhanced_personalization}), which improves upon the baseline by incorporating personalization. 


\subsubsection{Semantic Model} ~\\
The results for our baseline model are presented in Table \ref{tab:recall_model_result}. This Non-personalized model (as detailed in Section \ref{subsec:baseline_model}) was tested on both dataset variants-NPER\_Test and PER\_Test, using \(\mathrm{Recall}@k\) with \(k \in \{8, 12, 24, 100\}\) as the evaluation metrics. The results indicate that, even without personalization, the semantic engine model remains effective and also surpasses the baseline models compared with. (DistilledBert\cite{Hofstaetter2021_tasb_dense_retrieval} and BM25\cite{trotman2014improvements}) 



\subsubsection{Enhanced Personalization Semantic Model} ~\\
Our enhanced personalization model (Section \ref{subsec:enhanced_personalization}) tailors search results using customer’s purchase history and personal attributes, falling back to a non-personalized approach when purchase history is unavailable. While some queries gain less from personalization (see Section \ref{personalization_effect_on_queries}), Table \ref{tab:recall_model_result} shows that our personalized model consistently outperforms the non-personalized model in Recall@K across all K (including 2–5). We focus on \(k=8\text{--}100\): \(k{=}8,12\) capture early, above-the-fold relevance; \(k{=}24\) approximates a full first page(the first product page); and \(k{=}100\) reflects multi-page exposure (first four pages) to reflect real‐world display constraints.  
Furthermore, by training on the combined personalized (PER\_Train) and non-personalized (NPER\_Train) datasets—and by inputting a zero vector when personalization data is missing—the model seamlessly blends personalized and non-personalized behavior, ensuring robust retrieval for both new and returning customers.





\subsubsection{Effectiveness of Taxonomy-Based Negative Sampling} ~\\
We evaluated four hard-negative sampling strategies, random, BM25-based \cite{karpukhin2020dense}, ANCE \cite{xiong2020approximatenearestneighbornegative}, and our taxonomy-based method using \(\mathrm{Recall}@k\) for \(k\!\in\!\{8,12,24,100\}\). As shown in Table \ref{tab:compare-negative-sampling-techniques}, our taxonomy-based approach not only delivers the highest retrieval quality (\(63.50/69.88/77.17/83.66\%\) at \(k{=}8/12/24/100\), outperforming random \(41.89/44.53/47.97/51.96\%\), BM25 \(38.37/40.89/44.25/48.97\%\), and ANCE \(45.96/48.43/52.32/60.43\%\)), but also reduces sampling overhead. By restricting the search for negatives to a small, hierarchically related subtree of the catalog, rather than scanning the entire index or re-encoding huge corpora, our method cuts down data-preparation time by orders of magnitude while generating more contextually challenging negatives.

\subsubsection{Personalization Effects 
on Different Queries} \label{personalization_effect_on_queries} ~\\
We assessed our personalization strategy using two dimensions: \emph{query specificity} and \emph{query frequency}. Query specificity distinguishes between specific queries (narrowly defined, e.g., “white ceramic bathroom sink”) and general queries (broader, e.g., “bathroom fixtures”). Specificity is determined by the normalized purchase-entropy of a query’s item distribution (Ai et al. \cite{ai2019zero}), with low-entropy queries classified as specific and high-entropy queries as general. 

Query frequency segments head queries (high-volume, top-percentile searches) from tail queries (infrequent, bottom-percentile, long-tail searches). While prior research suggests selective personalization based on query type, our model consistently achieves significant improvements across all query segments.


As shown in Table~\ref{tab:selective_personalization}, personalization improves \(\mathrm{Recall}@k\) for all query segments and cutoffs \(k\!\in\!\{8,12,24,100\}\). The largest early-rank lifts are for \emph{Specific} queries \((+98.68\%@8,\ +69.88\%@12,\ +39.06\%@24,\ +11.93\%@100)\), with \emph{General} queries close behind \((+96.65/+69.71/+35.15/+8.37)\). \emph{Head} queries gain \((+77.94/+60.64/+26.24/+7.78)\). At deeper ranks, \emph{Tail} queries see the biggest improvement at \(k{=}100\) \((+12.50\%)\) while also rising \((+98.24/+69.30/+16.03)\) at \(k{=}8/12/24\). Overall, personalization consistently outperforms the non-personalized baseline, with the strongest relative gains on intent-rich Specific queries and meaningful benefits for long-tail retrieval.

\begin{table*}[t]
\caption{Impact of our personalization model on different query types, comparing high-frequency (Head) versus low-frequency (Tail) searches and narrowly focused (Specific) versus broadly phrased (General) queries. Cells report Recall@$k$ for $k\!\in\!\{8,12,24,100\}$; the green values in parentheses are relative lifts over the non-personalized baseline. 
Personalization improves recall for every segment, with the largest early-rank gains on \emph{Specific} queries (e.g., $+98.68\%$ at $k{=}8$) and the largest deep-rank gain on \emph{Tail} queries at $k{=}100$ ($+12.50\%$), indicating strong benefits on intent-rich and long-tail retrieval alike.}
\centering
\small
\begin{tabular}{llrrrrr}
\hline
\textbf{Model Type} & \textbf{Segmentation} &\textbf{Recall@8} &\textbf{Recall@12} & \textbf{Recall@24} & \textbf{Recall@100} \\ 
\hline
\multirow{4}{*}{w/o Personalization} 
& Specific & 38.77 & 47.75 & 62.59 & 80.65  \\
& General & 40.04 & 48.59 & 65.03 & 85.98 \\
& Head & 45.33 & 54.16 & 70.5 & 90.00  \\
& Tail & 41.46 & 52.19 & 60.76 & 80.00  \\
\hline
\multirow{4}{*}{w/ Personalization} 
& Specific & 77.03 \p{+98.68\%} & 81.12 \p{+69.88\%} & 87.04 \p{+39.06\%} & 90.27\p{+11.93\%} \\
& General & 78.74\p{+96.65\%} & 82.46\p{+69.71\%} & 87.89\p{+35.15\%} & 93.18\p{+8.37\%} \\
& Head & 80.66\p{+77.94\%} & 87\p{+60.64\%} & 89.00\p{+26.24\%} & 97.00\p{+7.78\%}  \\
& Tail & 82.19\p{+98.24\%} & 88.36\p{+69.30\%} & 70.50\p{+16.03\%} & 90.00\p{+12.5\%}  \\
\hline
\end{tabular}

\label{tab:selective_personalization}
\end{table*}



\subsection{Online Evaluation Result}
Our online evaluation was carried out via A/B testing in a live production environment. In this evaluation, a portion of live traffic was routed to our semantic model while the remaining traffic continued to use our legacy Customer Behavior and Lexical Matching models.  This online evaluation allows us to measure real-world impact of our model. 

\subsubsection{Impact On Business Metrics} ~\\
Quantitatively, we observe a 2.70\% increase in Conversion Rate(CR), 2.04\% gain in total visits resulting in items Added To Cart(ATC), and a 0.6\% increase in Average Order per Visit(AOV). We also evaluated the model on gross demand and expected revenue with improvements on these metrics as well. Due to the internal policy, we couldn't disclose the exact gross demand and revenue from the new models. 


\subsubsection{Retrieval and Latency Evaluation} ~\\
During A/B testing in production, our model met all Service Level Objectives targets for retrieval via our internal ANN service. End-to-end latency is primarily driven by real-time embedding generation (optimized to $\sim 50$ ms at the 95th percentile) and Google ScaNN-powered ANN search ($\sim 5$ ms at the 95th percentile).

\subsection{Analysis}
\subsubsection{Semantic Embeddings Enhance Cold-Start Items Retrieval}~\\
\emph{Please note that we have redacted the names of items as per internal policy}. Using the query “Power Drill”, our model surfaces "Items B and C", newly added with minimal interaction yet strong semantic relevance, over "Items D, E, and F," which, despite rich historical interactions, are less relevant.

As presented in Table \ref{tab:cold-start-comparison}, the legacy system overranks "Items D, E, and F", despite their low relevance, because of rich interaction histories, and underranks the more relevant, low‐interaction "Items B and C". Our Semantic Engine flips this: it assigns higher similarity scores to "Items B and C" and lower scores to "Items D, E, F", correctly prioritizing semantic relevance even when interaction data is minimal.



\begin{table}[t]
\caption{
Comparison of model performance for the query “Power Drill.” "Items B and C" have low interaction counts yet high semantic relevance, whereas "Items D, E, and F" exhibit strong interaction histories but low relevance. The legacy system overranks "Items D, E, and F" based on historical signals (e.g., purchases, cart additions), while our Semantic Engine prioritizes the truly relevant cold-start items (B and C), thus effectively addressing the cold-start problem.
}
\centering
\small
\begin{tabular}{llrrr}
\hline
\textbf{Relevance} & \textbf{Item} & \textbf{Interaction \#} & \textbf{Old Score} & \textbf{Sem. Score}\\ 
\hline
\multirow{3}{*}{Low} 
& D & 258,227 & 0.8355 &0.4396  \\
& E & 348,189 &0.9242 &0.4453 \\
& F & 101,498 &0.5613 &0.4403  \\
\hline
\multirow{2}{*}{High} 
& B & 158 & 0.2126 &0.6903  \\
& C & 132 & 0.2071 &0.6679 \\
\hline
\end{tabular}
\label{tab:cold-start-comparison}
\end{table}

\subsubsection{Frequent Vs Infrequent Shoppers Query} 
\begin{figure*}[!ht]
    \centering
\includegraphics[width=.55\linewidth]{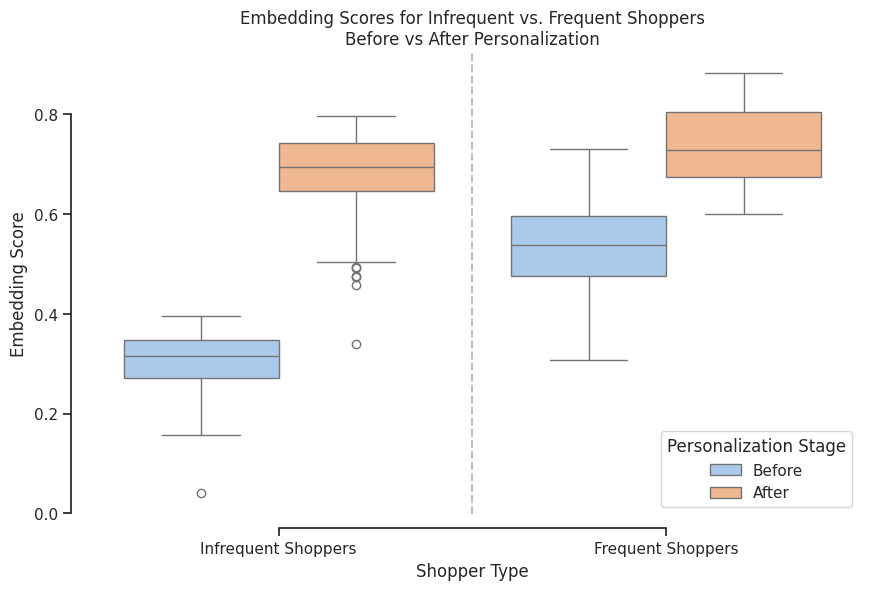}
    \caption{Embedding Similarity scores for infrequent and frequent shopper queries before and after personalization. Higher scores indicate better alignment with relevant items. The results demonstrate that personalization improves search relevance by increasing embedding similarity scores across both shopper types. After personalization, both distributions shift upward, with a larger median lift for infrequent shoppers; the lower tail and outliers shrink and the interquartile range narrows, indicating more stable relevance. Frequent shoppers begin higher and still gain (refinement), while the gap between the two cohorts narrows (correction of underspecified queries). Overall, personalization recovers missing brand/attribute cues for broad queries and sharpens already‐specific ones.}
    \label{fig:boxplot-embedding-scores}
\end{figure*}

~\\
This analysis highlights that infrequent shoppers tend to submit broad, underspecified queries (e.g., “carpet”), while frequent shoppers use more detailed or branded terms (e.g., “hearthstone inner peace carpet” or “duralux cruiser blue”), often leveraging synonyms or feature names to improve relevance. Such behavior gaps underscore the value of synonym mapping and query refinement for infrequent users. Figure \ref{fig:boxplot-embedding-scores} compares embedding similarity scores for both groups before and after our Personalized Semantic Engine, showing that personalization markedly improves query–item alignment across shopper segments.




\subsubsection{Significance Test Showing Effect of Personalization on Score and Distance} ~\\
\noindent Personalization significantly increased embedding similarity for both frequent and infrequent shoppers (paired $t = 13.22$, $p = 0.0002$), aligning results more closely with shopper intent and reducing query–item embedding distances. Moreover, it narrowed the performance gap between shopper types, enhancing search quality across all segments. 


\subsubsection{Generalization of Taxonomy-Based Hard Negative Sampling}
To assess generalizability, we evaluated our taxonomy-based hard negative sampling on a second, out-of-domain dataset: the public \textit{Amazon ESCI} corpus. We kept the training and evaluation protocols fixed and compared taxonomy-based hard negative against three common alternatives-BM25 \cite{karpukhin2020dense}, ANCE \cite{xiong2020approximatenearestneighbornegative}, and random negative sampling \cite{breiman2001random}. As summarized in Table~\ref{tab:negative-sampling-techniques-generalization}, Taxonomy-based hard negative sampling consistently outperforms these baselines across the Recall@k metrics (k=8, 12, 24, 100), demonstrating that the benefits of our sampling strategy transfer beyond the home-improvement domain to broader e-commerce categories.

\begin{table*}[t]
\caption{Recall performance comparison of sampling techniques to show the \textbf{generalization} of our Taxonomy-based sampling technique using the Amazon ESCI dataset. Taxonomy-based negative sampling significantly outperforms other sampling methods in Recall@K = (8,12,24,100) metrics.}
\centering
\small
\begin{tabular}{lrrrr}
\hline
\textbf{Sampling Technique} & \textbf{Recall@8} & \textbf{Recall@12} & \textbf{Recall@24} & \textbf{Recall@100} \\
\hline
Random Negative \cite{breiman2001random} & 19.15 & 22.68 & 28.93 & 40.08 \\
Karpukhin et al (BM25) \cite{karpukhin2020dense}  & 24.41 & 28.66 & 35.42 & 45.60 \\
Xiong et al (ANCE) \cite{xiong2020approximatenearestneighbornegative} & 22.18 & 34.14  & 43.76 & 54.21 \\
Taxonomy-based Negative (Ours) & \textbf{28.64} & \textbf{34.19} & \textbf{45.69} & \textbf{61.12} \\
\hline
\end{tabular}
\label{tab:negative-sampling-techniques-generalization}
\vspace{-.30cm}
\end{table*}

\subsubsection{Computational Efficiency and Latency of Taxonomy-Based Hard Negative Sampling Technique}
\label{subsec:tb-hns-efficiency}
Let $N$ be the catalog size and let $\mathcal{C}_p(Q)$ denote the sibling set under the parent category of the query item $Q$ in Algorithm~\ref{alg:taxonomy_hard_negative_sampling}. Our taxonomy-based hard negative sampling (TB-HNS) draws from $\mathcal{C}_p(Q)$ only (Steps 2–6), avoiding corpus-wide scans.

\paragraph*{Time complexity.}
When we precompute a map $\texttt{parent\_id}\!\to\!\texttt{[item ids]}$ and maintain $P$ as a hash set, a draw+membership check (Steps 6–8) is $O(1)$. Let
\[
\rho \;=\; \frac{|P \cap \mathcal{C}_p(Q)|}{|\mathcal{C}_p(Q)|}
\]
be the fraction of candidates that are positives. The expected number of trials in the rejection loop (Steps 5–11) is $1/(1-\rho)$, giving
\[
\mathbb{E}[T_{\text{TB-HNS}}] \;=\; O\!\big(1/(1-\rho)\big).
\]
In typical retail taxonomies $|\mathcal{C}_p(Q)| \!\ll\! N$ and $\rho$ is small, so TB-HNS is effectively $O(1)$ per negative. The worst case (when $\rho\!\approx\!1$) is bounded by $O(K)$ and returns \texttt{None} (Step 12).

\paragraph*{Latency versus baselines.}
\emph{BM25} requires an inverted-index query over $N$ items and materializes a top-$k$ list per request; \emph{ANCE} requires ANN probing over the embedding index and periodic re-encoding of the corpus. Both incur non-trivial per-query latency and maintenance costs. Our TB-HNS reduces mining to (i) one parent-lookup, (ii) one constant-time sample from the small array $\mathcal{C}_p(Q)$ (alias sampling optional), and (iii) one hash lookup against $P$. This removes global index scans and GPU forward passes, yielding substantially lower end-to-end data-prep time while keeping (and in our experiments improving) retrieval quality.

\subsubsection{Taxonomy Construction and Validation}
Our taxonomy is derived from the large retailer’s production catalog, which organizes items into a hierarchical (root$\rightarrow$leaf) structure; each product is mapped to a single path in this hierarchy. We validate it via random spot checks to ensure SKU$\rightarrow$category alignment and by comparing negative-selection scopes- in-batch random vs.\ taxonomy-based at the parent and grandparent levels. To prevent leakage, each product is canonicalized to a single taxonomy path based on its primary category, with duplicates removed during sampling. These checks confirm that the taxonomy reflects real product relationships and thereby strengthens our negative sampling strategy (Algorithm~\ref{alg:taxonomy_hard_negative_sampling}).

\section{Conclusion}
In this work, we’ve developed a semantic retrieval engine for e-commerce that combines a novel taxonomy-based hard-negative sampling strategy with user personalization to strengthen distinctions between closely related items and tailor results to individual preferences. Offline evaluations show clear recall gains over baselines, and live A/B testing delivers notable improvements in add-to-cart rates, average order values, and overall conversion.  



\bibliographystyle{IEEEtran}   
\bibliography{IEEE-Submission/sample-base}   

\vspace{12pt}

\end{document}